# INJECTION AND IMAGING OF ACHIRAL MICROSWIMMERS IN ZEBRAFISH

**Biao Wang**
Department of Mechanical and Energy Engineering
Southern University of Science and Technology
Shenzhen, China 518055
1369728199@qq.com

**U Kei Cheang***
Department of Mechanical and Energy Engineering
Southern University of Science and Technology
Shenzhen, China 518055
cheanguk@sustech.edu.cn

## ABSTRACT

Achieving stable *in vivo* locomotion is essential for using magnetically actuated microswimmers for biomedical applications; however, while existing microswimmers have excellent motion control *in vitro*, their motion is greatly hindered inside living organisms. Moreover, previous work had only visually demonstrated *in vivo* motion through gradient pulling or rolling, but not swimming. This study investigated the injection and imaging of the achiral planar microswimmers (APMs) inside a live zebrafish embryo. The APMs can be actuated under a rotating magnetic field to generate a forward thrust at low Reynolds number environment. Combined with a safe injection technique and clear *in vivo* imaging at high resolution, it would be possible to control the swimming motion of APMs inside the zebrafish embryo. This work shows the safe injection and the clear imaging of an APM in a transparent zebrafish model, demonstrating the possibility for follow-up in-depth studies of the swimming motion of microswimmers *in vivo*.

## 1   Introduction

Magnetically actuated microswimmers that can swim in bulk fluid, such as helical microswimmers (Dong et al., 2022), have garnered considerable attention due to their effective propulsion at low Reynolds number, precise targeting in complex environments, and potential for biomedical applications, such as drug delivery, cell transport, microsurgical procedures, and microdevice imaging (Li et al., 2018, Aziz et al., 2020, Xie et al., 2020, Song et al., 2022, Chen et al., 2023, Nelson and Pané, 2023). While there are many ways to actuate microrobots (Ghosh and Fischer, 2009, Ahmed et al., 2016, Karshalev et al., 2018), magnetic actuation can achieve precise motion control and is relatively safe (Yang and Zhang, 2020). Under magnetic actuation, microrobots can be driven to rotate using a rotating magnetic field (RMF) (Huang et al., 2015, Wang et al., 2018, Magdanz et al., 2020, Wu et al., 2022, Christiansen et al., 2023, Su et al., 2023). The rotational motion can be converted into translational swimming motion(Purcell, 1977). While previous works (Barbot et al., 2016, Hu et al., 2018)demonstrate the exceptional capability of microrobots to overcome different environments and situations *in vitro*, such as navigating geometrically complex pathways, non-Newtonian fluids, and biomimetic environments, few works were able to clearly show discernible motion of microrobots in *in vivo* environments.

Within the past decade, several works demonstrated motion control *in vivo* using mouse models (Yan et al., 2017, Xing et al., 2021, Song et al., 2022, Wrede et al., 2022). Mouse models are ideal for testing the overall therapeutic function of microrobots, but are a poor model to observe the motion of microrobots due to being opaque. Zebrafish models, on the other hand, are transparent; thus, they are ideal models to observe the motion of microrobots *in vivo*. Li *et al*. and Wu *et al*. demonstrated the control of spherical microrobots using a gradient field and an RMF, respectively, in the yolk of a zebrafish embryo. In both cases, the motion of individual microrobots was visually observed, where the motion type of the former was magnetic pulling (Li et al., 2018) and the latter was rolling (Wu et al., 2022). In the work by Wu *et al.,* the rotation of the microrobot was not discernible due to the difficulty in seeing the rotation of a sphere under limited imaging resolution. Among all of the studies involving microrobots in zebrafish, none showed the steady swimming motion of RMF-actuated microswimmers, helical or otherwise, with discernible rotational and translational motion.

---

* Corresponding author: U Kei Cheang (cheanguk@sustech.edu.cn)

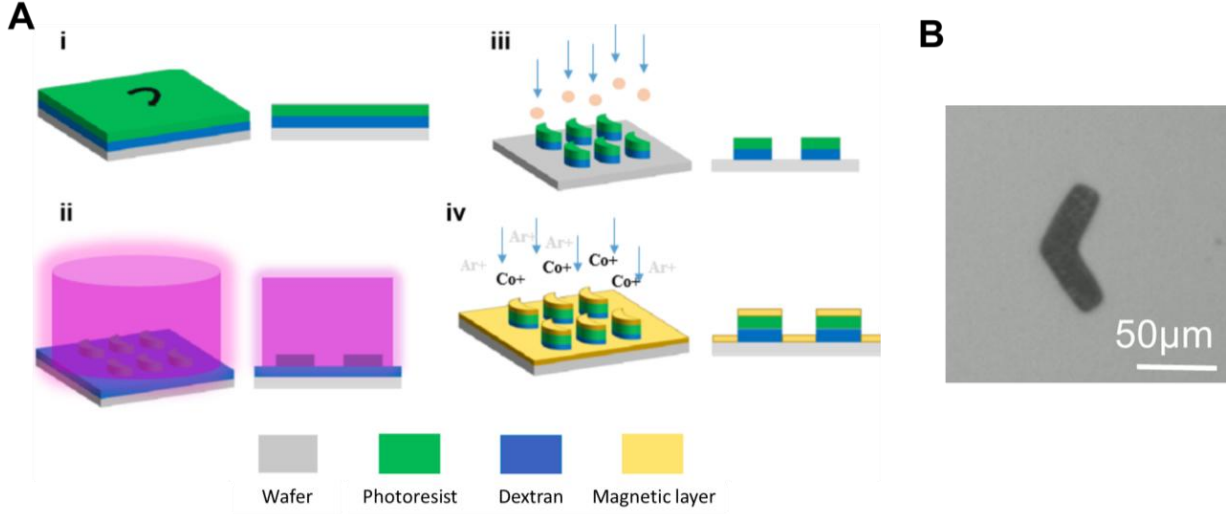

Figure 1: Fabrication of APMs. (A) Fabrication process of the APMs: i) spincoating, ii) UV exposure, iii) plasma treatment, and iv) thin film coating. (B) Microscopy image of a fabricated APM.

In this work, we fabricated APMs and demonstrated the possibility of imaging the APMs inside living zebrafish models. The design of the APMs was inspired by previous reports on achiral microswimmers (Tan et al., 2022). Through experiments, the APMs were clearly imaged inside a live zebrafish embryo using a conventional brightfield microscope, where their structures were visible with high resolution. The results shown verified the feasibility of studying swimming motion inside living organisms and will facilitate future in-depth studies on steady swimming *in vivo*, which will eventually enhance our understanding of using microswimmers for *in vivo* applications and address the age-old challenge of controlling microswimmers to swim inside the body.

## 2 Results

### 2.1 Fabrication of APMs

The dimensions of the APMs are as follows: length 50 μm, width 20 μm, and height 5 μm (with a width-to-height ratio of 4:1). The angle between the two arms is 120°. These dimensions were theoretically investigated to have optimal efficiency (Tan et al., 2022). The magnetic layer consists of a 300 nm thick Co layer, with an additional 15 nm of Ti deposited on top to prevent oxidation of the magnetic layer and ensure biocompatibility for *in vivo* experiments. The fabrication process is shown in Figure 1A. A microscope image of an APM is shown in Figure 1B.

### 2.2 Magnet control of APMs

The coils used in experiments are a Helmholtz coil system (Figure 2). For the microswimmers to be effectively controlled, we used a conical RMF (a RMF superimposed on a static magnetic field). The conical RMF **B** is expressed as:

$$\mathbf{B} = \begin{bmatrix} -B_s \cos(\theta_s) + B_r \sin(\theta_s) \cos(\omega t) \\ B_s \sin(\theta_s) + B_r \cos(\theta_s) \cos(\omega t) \\ B_r \sin(\omega t) \end{bmatrix} \quad (1)$$

where $B_r$ is the maximum amplitude of the RMF, $B_s$ is the magnitude of the static magnetic field, $\theta_s$ is the direction of rotation of the field, and $t$ is time. The direction of swimming is controlled by modulating the heading angle $\theta_s$.

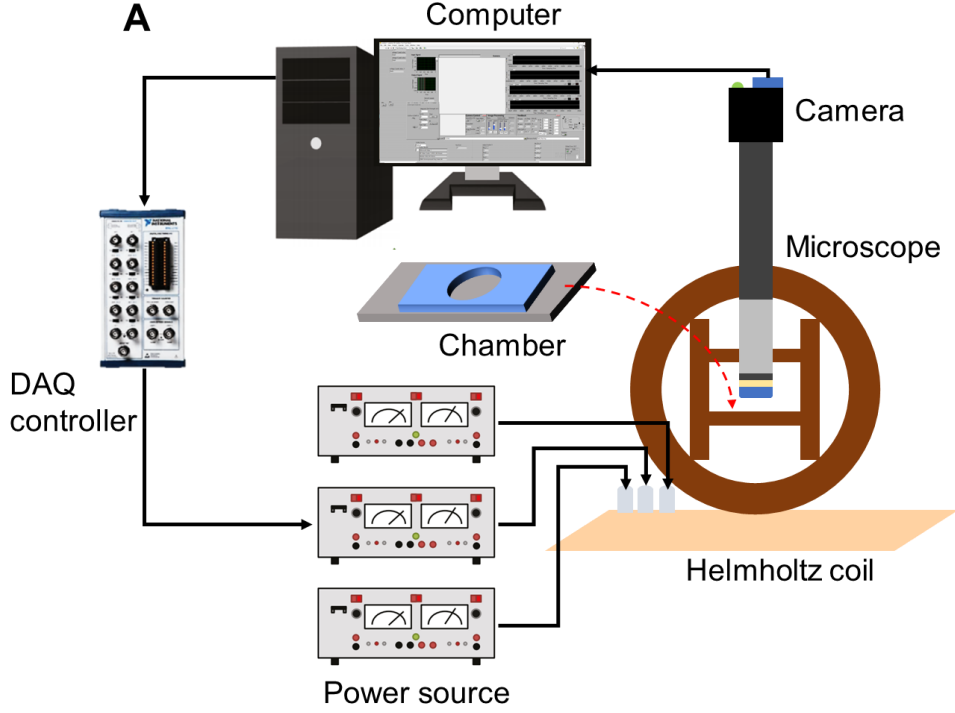

Figure 2: Magnetic control system with 3D Helmholtz coils.

The perpendicular vector to the plane of the rotating field can be expressed as:

$$\mathbf{n} = [-\cos(\theta_s) \ \sin(\theta_s) \ 0]^{\mathrm{T}} \tag{2}$$

In experiments, we modulated the motion speed of microswimmers by controlling the frequency of the RMF, and the motion direction by adjusting the angle of $\theta_s$. The magnetic field strength is calculated using a modified version of the Biot-Savart law, which is given as:

$$B_{coil} = \frac{\mu_0 n I R^2}{2(R^2 + x^2 - 2dx + d^2)^{\frac{3}{2}}} + \frac{\mu_0 n I R^2}{2(R^2 + x^2 + 2dx + d^2)^{\frac{3}{2}}} \tag{3}$$

where $\mu_0$ is the permeability, $n$ is the number of turns of wires per coil, $I$ is the electrical current passing through the wires, $R$ is the effective radius of the coil, $d$ is the distance between the coils, and $x$ is the coil distance to a point.

### 2.3  APM in Zebrafish Embryos

*In vivo* imaging was conducted in the yolk of the zebrafish embryo 28 hours post-fertilization. Zebrafish are widely used as animal models due to their genetic similarity to humans. A microprobe was used to create a wound on the anesthetized zebrafish embryo, and then the APM was pressed into the yolk. *In vivo* imaging experiments were conducted on the microswimmers using the brightfield microscope of the magnetic control system. This test is important because it is necessary to verify the feasibility of observing motion control *in vivo*, and to validate the imaging quality of the APMs inside the yolk using the imaging system of the magnetic control system. During the experiment, the magnification was set to 5×. In the representative result, the microswimmer was clearly imaged with a high definition resolution of 1280×1024 (Figure 3). The fertilized zebrafish embryos exhibited clear heartbeat observations, and the zebrafish remained alive throughout the experiment. These results demonstrated that it is possible to safely inject APMs into a live zebrafish embryo and to image them using the imaging system of the magnetic control system at high resolution.

## 3  Conclusion

In conclusion, this study successfully demonstrated the clear imaging of APMs within a live zebrafish embryo. While previous research had primarily focused on *in vitro* experiments, this investigation provided the visual confirmation of successful injection of APMs in zebrafish models at high resolution. Integrating injection and imaging techniques in this work with a magnetic control system capable of actuating APMs using an RMF enables the possibility of conducting future in-depth studies on the motion control of microswimmers *in vivo* with high-resolution imaging. This approach highlights the significant potential for advancing *in vivo* biomedical applications using magnetic microswimmers.

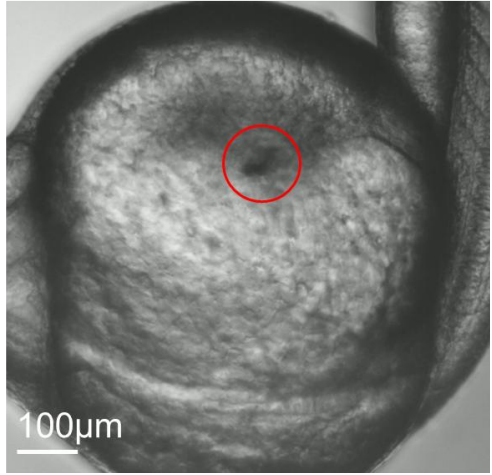

Figure 3: Imaging of an APM in the yolk of a zebrafish embryo.

## 4 Methods

### 4.1 APM Fabrication

APMs are fabricated using standard photolithography. First, a 5% w/v dextran solution is spin-coated onto a clean silicon wafer as a water-soluble sacrificial layer. Then, SU8-2005 photoresist is spin-coated at 2300 rpm for 30 s to obtain a 5 μm coating. Second, the sample is placed on a hotplate at 95 °C for 150 s pre-baking, and then placed at room temperature until the photoresist cools and hardens. The mask is then brought into hard contact with the photoresist and exposed to ultraviolet light for a specific time. The sample is then placed on a hotplate at 95 °C for 180 s post-baking, and the APM structure can be clearly seen on the silicon wafer at the same time. Third, the sample is placed in a developer solution to fully develop the exposed area, removing the unexposed part. The developed sample is then subjected to plasma treatment, which removes the excess dextran from the silicon wafer. Finally, a Ti-Co-Ti coating is deposited on the microswimmer using physical vapor deposition (PVD), with thicknesses of 15 nm, 300 nm, and 15 nm, respectively. The bottom layer of titanium is to enhance the adhesion to the photoresist, the middle layer of cobalt is the magnetic layer, and the top layer of titanium is the biocompatible layer. The sample is then placed in water, since only the dextran below the microswimmers is left, the microswimmers can be released in water.

### 4.2 Magnetic Actuation, Data Acquisition, and Analysis

In the experiment, a conical RMF generated by a three-dimensional Helmholtz coil was used to drive the microswimmers. An optical microscope system with a 5× objective lens and a CMOS camera (BFS-U3-13Y3M-C) was mounted on the coil system to observe and record the microswimmers. The coil system consisted of three pairs of electromagnetic coils, which were powered by three power supplies (Kepco, BOP20-5M) and controlled by a data acquisition (DAQ) device (National Instruments, PCI-6259). LabVIEW was used as the control window to control the DAQ devices and the camera. In the experiment, the microswimmers were transferred to a PDMS chamber using a micropipette.

### 4.3 Injection of APMs into the Yolk of Zebrafish Embryo

Zebrafish embryos used in the experiment were 28 hours post-fertilization. Our experiments showed that embryos less than 20 hours post-fertilization would lead to a large outflow of yolk during injection, resulting in the death of the zebrafish embryos. First, the zebrafish embryos were placed in a PDMS chamber filled with E2 culture medium.

Second, the outermost layer of the zebrafish embryo membrane was torn off using fine tweezers, and 40 μl of anesthetic was added. The zebrafish embryos were then moved to the edge of the chamber for injection. Third, a microprobe with a diameter of about 30 μm was used to make a wound in the yolk. Last, the microprobe was used to move the microswimmer to the wound, and then press the microswimmer into the wound.

## Acknowledgements

This work is funded by the Science, Technology, and Innovation Commission of Shenzhen Municipality (RCYX20210609103644015).

## References

Yue Dong, Lu Wang, Veronica Iacovacci, Xiaopu Wang, Li Zhang, and Bradley J Nelson. Magnetic helical micro-/nanomachines: Recent progress and perspective. *Matter*, 5(1):77-109, 2022.

Junyang Li, Xiaojian Li, Tao Luo, Ran Wang, Chichi Liu, Shuxun Chen, Dongfang Li, Jianbo Yue, Shuk-han Cheng, and Dong Sun. Development of a magnetic microrobot for carrying and delivering targeted cells. *Science Robotics*, 3(19):eaat8829, 2018.

Azaam Aziz, Stefano Pane, Veronica Iacovacci, Nektarios Koukourakis, Jürgen Czarske, Arianna Menciassi, Mariana Medina-Sánchez, and Oliver G Schmidt. Medical imaging of microrobots: Toward in vivo applications. *ACS Nano*, 14(9):10865-10893, 2020.

Meihua Xie, Wei Zhang, Chengying Fan, Chu Wu, Qishuai Feng, Jiaojiao Wu, Yingze Li, Rui Gao, Zhenguang Li, and Qigang Wang. Bioinspired soft microrobots with precise magneto-collective control for microvascular thrombolysis. *Advanced Materials*, 32(26):2000366, 2020.

Xin Song, Rujie Sun, Richard Wang, Kun Zhou, Ruoxiao Xie, Junliang Lin, Dimitar Georgiev, Andrei-Alexandru Paraschiv, Ruibo Zhao, and Molly M Stevens. Puffball-Inspired Microrobotic Systems with Robust Payload, Strong Protection, and Targeted Locomotion for On-Demand Drug Delivery. *Advanced Materials*, 34(43):2204791, 2022.

Zhi Chen, Xiaoxia Song, Xueliang Mu, Junkai Zhang, and U Kei Cheang. 2D Magnetic Microswimmers for Targeted Cell Transport and 3D Cell Culture Structure Construction. *ACS Applied Materials & Interfaces*, 15(7):8840-8853, 2023.

Bradley J Nelson and Salvador Pané. Delivering drugs with microrobots. *Science*, 382(6675):1120-1122, 2023.

Ambarish Ghosh and Peer Fischer. Controlled propulsion of artificial magnetic nanostructured propellers. *Nano Letters*, 9(6):2243-2245, 2009.

Suzanne Ahmed, Wei Wang, Lanjun Bai, Dillon T Gentekos, Mauricio Hoyos, and Thomas E Mallouk. Density and shape effects in the acoustic propulsion of bimetallic nanorod motors. *ACS Nano*, 10(4):4763-4769, 2016.

Emil Karshalev, Berta Esteban-Fernández de Ávila, and Joseph Wang. Micromotors for "Chemistry-on-the-Fly". *Journal of the American Chemical Society*, 140(11):3810-3820, 2018.

Zhengxin Yang and Li Zhang. Magnetic actuation systems for miniature robots: A review. *Advanced Intelligent Systems*, 2(9):2000082, 2020.

Tian-Yun Huang, Mahmut Selman Sakar, Angelo Mao, Andrew J Petruska, Famin Qiu, Xue-Bo Chen, Stephen Kennedy, David Mooney, and Bradley J Nelson. 3D printed microtransporters: Compound micromachines for spatiotemporally controlled delivery of therapeutic agents. *Advanced Materials*, 27(42):6644, 2015.

Xiaopu Wang, Chengzhi Hu, Lukas Schurz, Carmela De Marco, Xiangzhong Chen, Salvador Pané, and Bradley J Nelson. Surface-chemistry-mediated control of individual magnetic helical microswimmers in a swarm. *ACS Nano*, 12(6):6210-6217, 2018.

Veronika Magdanz, Islam SM Khalil, Juliane Simmchen, Guilherme P Furtado, Sumit Mohanty, Johannes Gebauer, Haifeng Xu, Anke Klingner, Azaam Aziz, and Mariana Medina-Sánchez. IRONSperm: Sperm-templated soft magnetic microrobots. *Science Advances*, 6(28):eaba5855, 2020.

Zehao Wu, Yuting Zhang, Nana Ai, Haoran Chen, Wei Ge, and Qingsong Xu. Magnetic Mobile Microrobots for Upstream and Downstream Navigation in Biofluids with Variable Flow Rate. *Advanced Intelligent Systems*, 4(7):2100266, 2022.

Michael G Christiansen, Lucien R Stöcklin, Cameron Forbrigger, Shashaank Abhinav Venkatesh, and Simone Schuerle. Inductive sensing of magnetic microrobots under actuation by rotating magnetic fields. *PNAS*, 2(9):pgad297, 2023.

Lin Su, Dongdong Jin, Yuqiong Wang, Qinglong Wang, Chengfeng Pan, Shuai Jiang, Haojin Yang, Zhengxin Yang, Xin Wang, and Neng Xia. Modularized microrobot with lock-and-detachable modules for targeted cell delivery in bile duct. *Science Advances*, 9(50):eadj0883, 2023.

Edward M Purcell. Life at low Reynolds number. *American journal of physics*, 45(1):3-11, 1977.

Antoine Barbot, Dominique Decanini, and Gilgueng Hwang. On-chip microfluidic multimodal swimmer toward 3D navigation. *Scientific Reports*, 6(1):19041, 2016.

Wenqi Hu, Guo Zhan Lum, Massimo Mastrangeli, and Metin Sitti. Small-scale soft-bodied robot with multimodal locomotion. *Nature*, 554(7690):81-85, 2018.

Xiaohui Yan, Qi Zhou, Melissa Vincent, Yan Deng, Jiangfan Yu, Jianbin Xu, Tiantian Xu, Tao Tang, Liming Bian, and Yi-Xiang J Wang. Multifunctional biohybrid magnetite microrobots for imaging-guided therapy. *Science Robotics*, 2(12):eaaq1155, 2017.

Jiehua Xing, Ting Yin, Shuiming Li, Tiantian Xu, Aiqing Ma, Ze Chen, Yingmei Luo, Zhengyu Lai, Yingnian Lv, and Hong Pan. Sequential magneto-actuated and optics-triggered biomicrorobots for targeted cancer therapy. *Advanced Functional Materials*, 31(11):2008262, 2021.

Xiaoxia Song, Rongxin Qian, Tingting Li, Wei Fu, Lijun Fang, Yuzhen Cai, Heng Guo, Lei Xi, and U Kei Cheang. Imaging-Guided Biomimetic M1 Macrophage Membrane-Camouflaged Magnetic Nanorobots for Photothermal Immunotargeting Cancer Therapy. *ACS Applied Materials & Interfaces*, 14(51):56548-56559, 2022.

Paul Wrede, Oleksiy Degtyaruk, Sandeep Kumar Kalva, Xosé Luis Deán-Ben, Ugur Bozuyuk, Amirreza Aghakhani, Birgul Akolpoglu, Metin Sitti, and Daniel Razansky. Real-time 3D optoacoustic tracking of cell-sized magnetic microrobots circulating in the mouse brain vasculature. *Science Advances*, 8(19):eabm9132, 2022.

Liyuan Tan, Zihan Wang, Zhi Chen, Xiangcheng Shi, and U Kei Cheang. Improving Swimming Performance of Photolithography-Based Microswimmers Using Curvature Structures. *Micromachines*, 13(11):1965, 2022.